\documentstyle[12pt,epsfig]{article}
\def\be{\begin{eqnarray}}
\def\ee{\end{eqnarray}}
\def\nn{\nonumber}
\def\ds{\displaystyle}

\def\de{\partial}
\def\ve{\varepsilon}
\def\x{{\vec x}}
\def\v{{\vec v}}
\def\bnabla{\mbox{\boldmath$\nabla$}}
\def\w{\mbox{\boldmath$w$}}

\def\bbv{\mbox{\boldmath$B$}}
\def\av{\mbox{\boldmath$a$}}
\def\bv{\mbox{\boldmath$b$}}
\def\sigmav{\mbox{\boldmath$\sigma$}}
\def\Av{\mbox{\boldmath${\cal A}$}}
\def\Lone{\mbox{\boldmath$l_1$}}
\def\Ltwo{\mbox{\boldmath$l_2$}}
\def\Bv{\mbox{\boldmath${\cal B}$}}
\def\lb{{l_B}}
\def\H{{\cal H}}
\def\ex{\hat{\mbox{\boldmath$x$}}}
\def\ey{\hat{\mbox{\boldmath$y$}}}
\def\ez{\hat{\mbox{\boldmath$z$}}}
\def\ev{\mbox{\boldmath$e$}}
\def\pla{\mbox{\boldmath$\mu$}}
\def\path{\mbox{\boldmath$\gamma$}}
\def\Pz{\Pi^{(0)}}
\def\Pu{\Pi^{(1)}}
\def\Xz{X_{(0)}}
\def\Xu{X_{(1)}}
\begin{document}
\begin{flushright}
{\large{\sl UPRF-96-482}}
\end{flushright}
\vskip2.0cm
\begin{center}
{\LARGE\sf   Quantum Charged Spinning Particles in a Strong}\\
\vskip0.2cm
{\LARGE\sf   Magnetic Field (a Quantal Guiding Center Theory)}\\
\vskip1.7cm
{\Large P.\ Maraner}\\
\medskip              
{\large\sl Dipartimento di Fisica, Universit\`a di Parma,}\\ 
\smallskip
{\large\sl and INFN, Gruppo collegato di Parma,}\\
\smallskip
{\large\sl  Viale delle Scienze, 43100 Parma, Italy}
\end{center}
\vskip2.0cm

\begin{abstract}
 A {\sl quantal guiding center theory} allowing to systematically 
study the separation of the different time scale behaviours of 
a quantum charged spinning particle moving in an external inhomogeneous  
magnetic filed is presented. A suitable set of operators adapting to 
the canonical structure of the problem and generalizing the 
kinematical momenta and guiding center operators of a particle coupled
to a homogenous magnetic filed is constructed. The Pauli Hamiltonian 
rewrites in this way as a power series in the magnetic length $l_B=
\sqrt{\hbar c/eB}$ making the problem amenable to a perturbative analysis. 
The first two terms of the series are explicitly constructed. The effective 
adiabatic dynamics turns to be in coupling with a gauge filed and 
a scalar potential. The mechanism producing such magnetic-induced 
geometric-magnetism is investigated in some detail.
\end{abstract}
\vskip1.0cm
\noindent
{\tt short title}: {\sf A Quantal Guiding Center Theory}\\
\medskip
{\tt PACS}:  {\sf 03.65.-w, 02.40+m, 52.20.Dq, 02.90.+p}

\thispagestyle{empty}
\newpage
\vskip1.0cm
\noindent
{\large\sf \ }
\thispagestyle{empty}

\newpage
\section{Introduction}

 The motion of a charged particle in a strong inhomogeneous 
magnetic filed is a nontrivial problem displaying a variety 
of very interesting phenomena ranging from chaos to phase
anholonomy.
 Being of utmost importance in plasma physics, expecially in the 
study of magnetic confinement, the subject has been worked out
in great detail in classical mechanics  with special attention 
to phenomenological implications as well as to formal 
aspects. The canonical structure of the problem, in particular, 
has been deeply investigated only in a relatively recent 
time by R.G.\ Littlejohn \cite{Lj} revealing the appearance 
of geometry induced gauge structures in the adiabatic motion of 
classical charged particles.
Very few, on the other hand, is known about the behaviour of 
quantum particles in strong inhomogeneous magnetic fields,
the  reason being essentially that the techniques developed for classical 
mechanics do not easily generalize to the quantum context. 
Some work has been done for neutral spinning particles by M.V.\ Berry 
\cite{Be86}, Y.\ Aharonov \& A.\ Stern \cite{A&S92} and R.G.\ Littlejohn 
\& S.\ Weigert \cite{L&W93} in connections with geometrical phases, whereas 
a quantum treatment for charged spinning particles is still missing.
It is the purpose of this paper to
present what may be probably called a 
{\sl quantal guiding center theory} in which the coupling 
between the spin and spatial degrees of freedom of a quantum charged 
spinning particle moving in a strong inhomogeneous magnetic field is 
systematically taken into account. This allows to extend to the quantum 
domain the previous classical results. Our treatment, essentially
algebraic in nature, is a re-elaboration and---we believe---a simplification 
of the technique originally proposed by R.G.\ Littlejohn in classical 
mechanics. It is based on a 
different choice of non-canonical variables adapting to classical as well as 
quantum mechanics. Depending essentially on the canonical structure 
the method applies indistinctly to the classical and the quantum theory. 
We nevertheless focus on the quantum problem.

\begin{figure}[t]
\centerline{\mbox{\epsfig{file=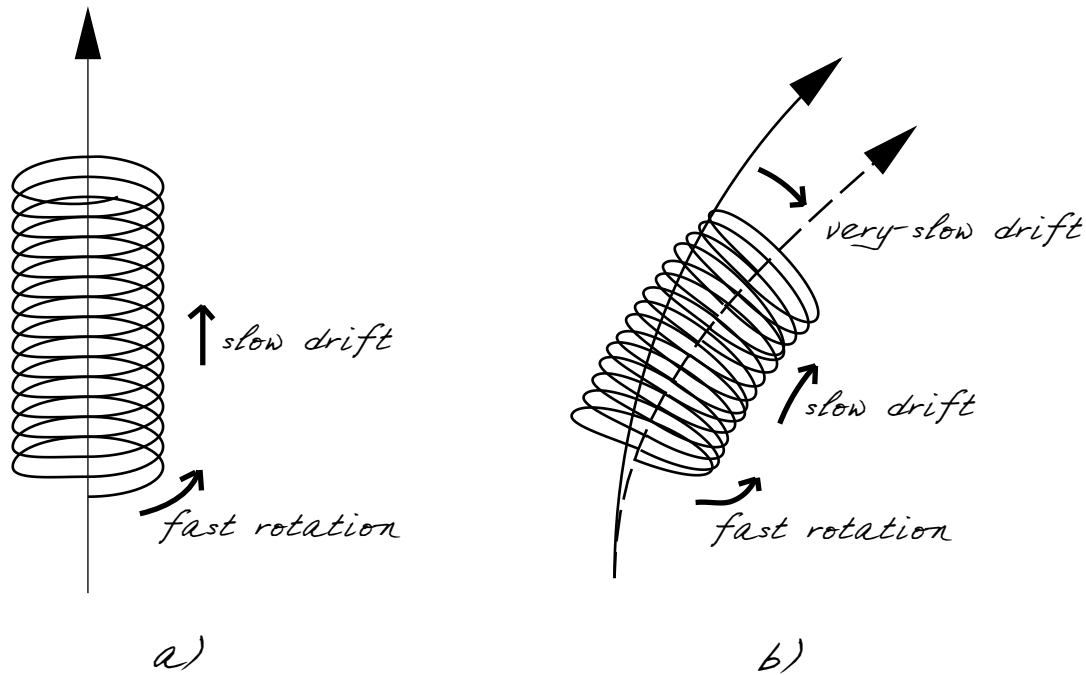}}}
\caption{Different time scale behaviours of a charged particle in
a strong magnetic field: {\sl a)} {\sl fast} rotation of the particle and 
{\sl slow} guiding center drift in a homogeneous field; {\sl b)} in
the inhomogeneous case the guiding center  drifts 
away from the filed line {\sl very-slowly}.}  
\label{fig1}
\end{figure}

In order to better understand what is going on in the  
strong-filed regime of a quantum particle moving in an external  magnetic 
filed it is better to first have in mind the main features of corresponding
the classical problem \cite{No63}.  Let us therefore to briefly consider 
a classical particle of mass $m$ and charge $e$ moving in a homogeneous 
magnetic filed of intensity $B$. As is well known the trajectory of the 
particle is represented by a spiral wrapping around a field line, as 
sketched in Fig.\ref{fig1}a: the particle performs a uniform circular motion 
of frequency $\omega_B=eB/mc$ and radius $r_B=mc|v_\perp|/eB$ ($|v_\perp|$ 
is the norm of the normal component of the velocity) in the directions normal 
to the field, while the center of 
the orbit, called the  {\sl guiding center}, moves freely along a filed line.
 Keeping fixed the initial condition, the stronger the magnetic field 
the faster  the rotation of the particle when compared 
with the drift along the filed direction and smaller the portion of space
explored by the particle around the filed line. This indicates, to the one
side,
the presence of different time scales in the dynamics of the system and
gives, on the other hand, the reason why the motion in a very strong magnetic 
filed may be studied along the same lines as that in a weakly inhomogeneous one.
 Let us introduce now a small inhomogeneity in the field.
The picture of the motion should not change substantially. The particle, 
in fact, keeps on rotating around its guiding center, the frequency and the
radius weakly depending now on the position, and the guiding center still
drifts along a field line. In this case, however, the guiding center do 
not remains exactly on a single field line, it starts drifting very-slowly 
in the directions normal to the filed.
 Three different time scale behaviours of the system may therefore be 
distinguished:
the {\sl fast rotation} of the particle around the guiding center,
the {\sl slow drift} of the guiding center along a magnetic field line and
the {\sl very-slow drift} of the guiding center in the direction normal to the
field. 
The situation is sketched in Fig.\ref{fig1}b. The stronger the 
magnetic field the cleaner  the separation among the three degrees of 
freedom.

 An outlook to the canonical structure of the homogeneous case 
makes immediately clear how the introduction of kinematical momenta
and  guiding center operators allows to separate the three
degrees of freedom of the system. This is briefly reported in section 2
where the relevant notations of the paper are also set up. 
After discussing the geometrical
complications involved in the adiabatic motion of a charged particle in 
an inhomogeneous magnetic field, section 3, an appropriate set of
non-canonical operators generalizing the one used in the discussion of 
the homogeneous problem is constructed in section 4. These are obtained as
formal power series in the magnetic length $l_B=\sqrt{\hbar c/eB}$
which appears naturally as the adiabatic parameter of the theory.
The Pauli Hamiltonian
describing the motion of the particle is then rewritten in terms of the new 
adiabatic operators in sections 5 and 6, whereas the anholonomic effects 
appearing in the adiabatic separation of the fast and slow/very-slow
degrees of freedom are discussed in section 7. Our results are summarized in 
equations (\ref{Hend}), (\ref{A}) and (\ref{V}). 
In the classical limit these reproduce correctly the classical theory.
Section 8 contains our conclusions.

\section{Canonical structure of the guiding center motion}

Magnetic interactions appearing essentially as modifications of 
the canonical structure of a dynamical system it is worthwhile 
to start by briefly discussing this peculiarity in the elementary
case of a quantum charged spinning particle in a homogeneous magnetic 
filed. This allows us to immediately focus on the heart of the problem
establishing at the same time terminology and notations. We consider 
therefore a spin-1/2 particle of mass $m$, charge $e$ and gyromagnetic 
factor $g$ moving in space under the influence of the {\em homogeneous} filed
$\bbv(\x)=B\,\ez$. As in the inhomogeneous case, to be discussed later on,
the physical dimension of the magnetic field is reabsorbed
in the scale factor $B$, the inverse square root of which, opportunely 
rescaled, will play the role of the adiabatic parameter of our theory.
 Introducing  an arbitrary choice of the vector potential $\av$ for 
the dimensionless filed $\bbv(\x)/B$, $\mbox{rot}\,\av=\ez$, the motion 
of the particle is described by the Pauli Hamiltonian
\be
\H={1\over2m}\Big(-i\hbar\bnabla -{eB\over c}\av\Big)^2
   +g{\hbar e B\over mc}\ez\cdot\sigmav
\label{H2}
\ee
$\bnabla=(\de_x,\de_y,\de_z)$ denoting the gradient operator and
$\sigmav=(\sigma_x,\sigma_y,\sigma_z)$ the matrix-valued vector
constructed by means of the three Pauli matrices. As well known the 
solution of this simple problem was first obtained by Landau at the 
beginning of the thirties and leads naturally to replace the 
standard set of canonical operators $p_i=-i\hbar\de_i$, $x^i$,
$i=1,2,3$, by  the gauge invariant {\sl kinematical momenta}
$\pi_i=p_i-(eB/c)a_i$ and the {\sl guiding center operators}
$X^i=x^i+(c/eB)\ve^{ij}\pi_j$. A very rapid computation yields
the nonvanishing commutation relation among the new variables
\begin{equation}
\begin{array}{ccc}
[\pi_2,\pi_1]=-i\ds{\hbar eB\over c},\hskip0.7truecm &
[\pi_3,  X^3]=-i\hbar,               \hskip0.7truecm &
[  X^1,  X^2]=-i\ds{\hbar c\over eB},
\end{array}
\label{cr2.1}
\end{equation}
indicating $\pi_2$-$\pi_1$, $\pi_3$-$X^3$ and $X^1$-$X^2$ as couples
of conjugates variables. Moreover, the scale dependence of the 
commutators (\ref{cr2.1}) allows to identify the three couple of 
operators as describing respectively the {\sl fast}, the {\sl slow}
and the {\sl very-slow} degrees of freedom of the system (see eg.\
\cite{Ma96}). In terms of the new gauge invariant operators
Hamiltonian (\ref{H2}) rewrites in the very simple form $\H=
({\pi_1}^2+{\pi_2}^2+{\pi_3}^2)/2m+g\hbar eB\sigma_3/mc$.
The harmonic oscillator term $({\pi_1}^2+{\pi_2}^2)/2m$ takes
into account the rapid rotation of the particle around its
guiding center while the free term ${\pi_3}^2/2m$ the slow drift of 
the guiding  center along the straight magnetic field lines. 
The very-slow variables $X^1$ and $X^2$ being constant 
of motion, the guiding center do not move in
the directions normal to the field. Let us  stress that
in the canonical formalism the spatial rotation of the particle around
its guiding center is taken into account by the phase space trajectory of a 
couple of conjugate variables: the particle's velocity components 
in the directions normal to the field: $\pi_1$ and $\pi_2$. 
The presence of an external
magnetic field produces therefore a rotation of the canonical structure,
mixing up spatial coordinates and canonical momenta in new canonical operators
adapting to the different time scale behaviours of the particle!
 In section 4 we will construct such ``adapted operators''---as power series 
in the adiabatic parameter---for the motion of a quantum charged spinning 
particle in an arbitrary magnetic filed. This allows to extend to quantum
mechanics the Hamiltonian approach to the classical guiding center motion 
developed by  R.G.\ Littlejohn \cite{Lj}.
The case of a magnetic filed with constant direction has been previously
considered in \cite{Ma96}. 
Before to proceed  some preparatory material is however necessary. 

First of 
all it  is convenient to introduce dimensionless quantities by factorizing 
the energy  scale $\hbar \omega_B$, $\omega_B=eB/mc$, from the Hamiltonian. 
This leads to  redefine kinematical momenta and guiding center operators as 
\be
& &\pi_i=-i\lb\de_i-\lb^{-1}a_i(\x) \label{km}\\
& &X^i=x^i+\lb\,\ve^{ij}\pi_j.    \label{gco}
\ee
$\lb=\sqrt{\hbar c/eB}$ being the {\sl magnetic length}. The relevant 
commutation relations may so be recast in the compact and very convenient 
form
\begin{equation}
\begin{array}{lcl}
\!\!
\begin{array}{l}
[\pi_i,\pi_j]=\,i\ve_{ij}\cr
[\sigma_i,\sigma_j]=\,i\ve_{ijk}\sigma_k
\end{array}                               \mbox{\Huge\}}& &\mbox{{\sl fast}}\cr
[\pi_i,X^j]=\,-i\lb \delta_i^3\delta_3^j  & &\mbox{{\sl slow}}\cr
[X^i,X^j]=\,-i\lb^2\ve^{ij}               & &\mbox{{\sl very-slow}}
\end{array}
\label{cr2.2}
\end{equation}
where the spin variables have also been considered. 

As a second and 
more serious task the geometrical structure responsible for the anholonomic
effects appearing in the adiabatic motion in a strong magnetic field has
to be discussed.

\section{Magnetism and geometric-magnetism}
 
 The beautiful analysis of the adiabatic separation of {\sl fast}
and {\sl slow} degrees of freedom in a quantum system proposed by 
M.V.\ Berry \cite{Be84-89}, H.\ Kuratsuji \& S.\ Iida \cite{K&I85},
 J.\ Moody, A.\ Shapere \& F.\ Wilczek \cite{M&S&W86}, R.\ Jackiw \cite{Ja88} 
and others,
has pointed out that in lowest order the reaction of the fast to 
the slow dynamics is through a geometry-induced gauge structure 
resembling that of (electro-)magnetism. This phenomenon has been
identified and found to be important in a variety of physical contexts
\cite{S&W89} and has been recently referred by M.V.\ Berry \& J.M.\ 
Robbins  as {\sl geometric-magnetism} \cite{B&R93}. 
A rather curious fact, first pointed out by R.G.\ Littlejohn in a series 
of beautiful papers on the canonical structure of classical  guiding center 
motion \cite{Lj}, is that, in some circumstances, magnetism 
itself may generate geometric-magnetism. The aim of the present section is 
that of discussing the geometry involved in such ``magnetic-induced 
geometric-magnetism''. 

\begin{figure}[t]
\centerline{\mbox{\epsfig{file=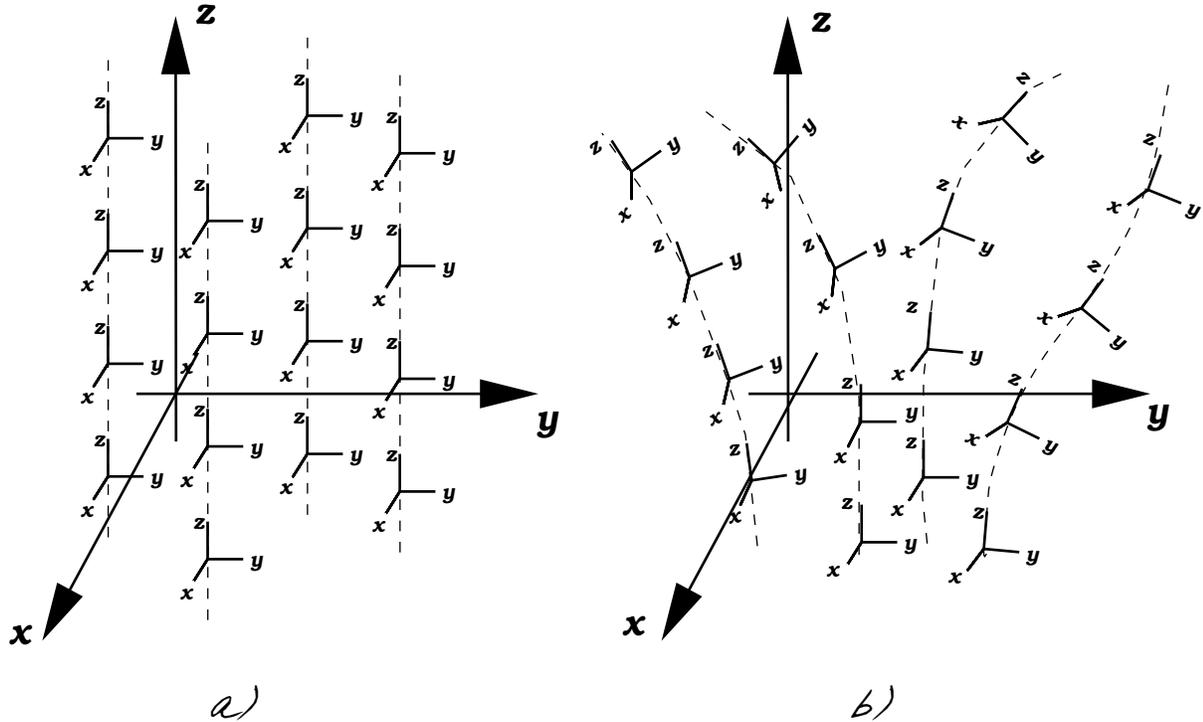}}}
\caption{Framing the tangent bundle $T\!R^3$ of the physical space: {\sl a)} 
by means of a single fixed frame in $R^3$; {\sl b)} by  using local reference 
frames adapting to the magnetic field lines geometry.}
\label{fig2}
\end{figure}

For shake of clearness it is useful to begin by briefly recalling 
the geometrical character of the kinematical quantities characterizing 
the motion of a particle in space. This will led to a rather intuitive 
picture of the geometrical structure involved in the adiabatic motion of 
a charged spinning particle in a strong magnetic field, allowing, at 
the same time, to frame it in a general and rigorous context. 
As well known the state of a particle moving in space is completely 
characterized by its position $\x$ and its velocity $\v$, i.e.\
by a point in the {\sl tangent bundle} $T\!R^3$ of the three-dimensional 
Euclidean space $R^3$. The flat parallel transport of $R^3$ makes it 
natural to parameterize every fiber $T_{\x}R^3$ of the bundle by means of a 
fixed reference frame in $R^3$, that is, to identify the tangent space in every 
point $\x$ with the physical space itself. Such an identification is 
certainly very useful in most circumstances, but it is a convention after 
all. In principle we are free to choose arbitrarily the frame of $T_{\x}R^3$ 
in every $\x$, the parallel transport---and not the way in which we describe 
it---being all that matters. This freedom of arbitrarily rotating the 
reference frame of the tangent space in every point $\x$, a local SO(3) 
symmetry, plays a crucial role in what follows. To visualize the situation, 
therefore, we shall picture the Euclidean space as filled up with orthonormal 
reference frames. To start with, we can imagine all them as combed 
parallel to a single fixed frame $\{\ex,\ey,\ez\}$ in $R^3$ 
(see Fig.\ref{fig2}a),
but even in a flat geometry, this is not always the better choice.

\subsection*{The magnetic line bundle} 

As qualitatively sketched above, the motion of 
a charged-spinning particle in a strong magnetic filed is characterized 
by the separation of time scales in the three degrees of freedom, making 
the system amenable to a perturbative analysis. In the lowest order 
approximation the particle performs a {\sl fast} rotation in the plane 
normal to the field line at which its guiding center is located. 
This is taken into account by the two components normal to the field 
of the particle's velocity (to this order a couple of conjugate variables). 
Disregarding the {\sl slow} drift of the guiding center along the filed 
line and the {\sl very-slow} motion, therefore, the velocity of a particle 
which guiding center is located in $\x$ is effectively constrained 
to the plane $\pla_{\x}$ generated by the vectors normal to the field
in $\x$. In every  point of the space the magnetic filed $\bv(\x)$ 
picks the complex line $\pla_{\x}$ out of the tangent space $T_{\x}R^3$,
reducing the {\sl tangent bundle} $T\!R^3$ to a complex line bundle,
hereafter the {\sl magnetic line bundle} ${\cal M}$.\footnote{This subbundle
of $T\!R^3$ may be identified with the {\sl plane bundle} of B.\ Felsager
\& J.M.\ Leinaas \cite{F&L80}. See also the related paper of F.\ Gliozzi 
\cite{Gl78}.}  
 It is then natural to use the local $SO(3)$ symmetry of the theory to adapt 
the parameterization of $T\!R^3$ to the subbundle ${\cal M}$ by combing, say, 
the $\ez$ direction of the frame of every $T_{\x}R^3$ according to the 
direction of the field. We so smoothly introduce point dependent adapted 
reference frames $\{\ev_1,\ev_2,\ev_3\}$ in such a way that in every point 
$\ev_1(\x)$, $\ev_2(\x)$ parameterize $\pla_{\x}$ while $\ev_3(\x)$ 
is aligned with $\bv(\x)$ (see Fig2b). Such reference frames are commonly 
used in the discussion of geometrically non trivial physical problems 
such as in general relativity and are  referred as {\sl anholonomic 
frames}. 
It is worthwhile to note that fixing  $\ev_3$ according to the 
filed direction reduces the local $SO(3)$ symmetry of $T\!R^3$
into the local $SO(2)\equiv U(1)$ symmetry of ${\cal M}$. 
The vectors $\ev_1(\x)$ and $\ev_2(\x)$ are  in fact determined up 
to the rotation
\begin{equation}
\begin{array}{lcr}
\ev_1(\x)&\rightarrow&\ev_1(\x)\,\cos\chi(\x)-\ev_2(\x)\,\sin\chi(\x)\\
\ev_2(\x)&\rightarrow&\ev_1(\x)\,\sin\chi(\x)+\ev_2(\x)\,\cos\chi(\x)
\end{array}
\label{ggt}
\end{equation}
$\chi(\x)$ being a point dependent angle. This residual ambiguity 
will result in the gauge freedom of our theory.

\subsection*{Magnetic line bundle geometry} 
We may now wonder how the vectors lying in ${\cal M}$ are transported 
from  point to point, that is, whether the geometry of the magnetic 
line bundle is trivial or not.  To this task we proceed in two steps.  
Considering a vector $\w(\x)=w^\nu\ev_\nu(\x)$, $\nu=1,2$, in $\pla_{\x}$,
we first transport it from the point $\x$ to the infinitesimally closest 
point $\x+d\x$ by means of the Euclidean parallel transport of $R^3$ and,
second, we project it onto the plane $\pla_{\x+d\x}$. (i)~the Euclidean 
parallel transport of $\w$ in $\x+d\x$ may be immediately evaluated as 
\be
\w(\x+d\x)=\,\w(\x)-w^\nu\,(\ev_\nu\cdot\de_k\ev_i)dx^k\, \ev_i,
\nn
\ee
latin indices running over $1,2,3$, greek indices over $1,2$ and where
the sum over repeated indices is implied\footnote{This notation will be
employed throughout the rest of this paper.}. The three 
quantities\footnote{The 
vectors $\ev_1$, $\ev_2$ and $\ev_3$ being orthonormal in every point $\x$, 
$\ev_i \cdot\ev_j=\delta_{ij}$, these are the only independent quantities.} 
$\ev_1\cdot \de_k\ev_2$, $\ev_1\cdot\de_k\ev_3$ and $\ev_2\cdot\de_k\ev_3$ 
characterize the flat parallel transport of $R^3$ in the anholonomic frame 
(it is in fact possible to make them vanishing by rotating the adapted frames 
$\{\ev_1 ,\ev_2,\ev_3\}$ back to fixed directions in every point). 
(ii)~the projection onto  $\pla_{\x+d\x}$ yields then
\be
\w(\x+d\x)|_{\pla}=\w(\x)- w^\mu\,(\ev_1\cdot\de_k\ev_2)dx^k
                           \ve_\mu^{\;\;\nu}\,\ev_\nu,
\nn
\ee
indicating that the parallel transport along the infinitesimal path 
connecting $\x$ to $\x+d\x$ produces the vector $\w$ to be rotated by the 
infinitesimal angle $d\alpha=(\ev_1\cdot\de_k\ev_2)dx^k$. When parallel 
transported along a finite closed path $\path$ the vector will therefore 
return to the starting point rotated by the angle \cite{F&L80}
\be
\alpha_{\path}=\oint_{\path} (\ev_1\cdot\de_k\ev_2)dx^k;
\nn
\ee
this quantity being in general different from zero, the geometry  of 
the magnetic line bundle results  not flat. The operation of locally 
projecting onto the plane $\pla$ reduces the trivial $SO(3)$ local 
symmetry of the theory to a non-trivial $SO(2)\equiv U(1)$ local 
symmetry! This local structure is described by a magnetic-like
$U(1)$ gauge theory. The parallel transport of the magnetic line bundle 
${\cal M}$, results  in fact completely characterized by the vector
\be
{\cal A}_k=\ev_1\cdot\de_k\ev_2,
\label{ggp}
\ee
the {\sl connection one-form} of ${\cal M}$. $\Av$  appears in 
the theory as a geometry-induced vector potential (to not be 
confused with the vector potential $\av$ representing the real 
magnetic field $\bv$). A point dependent redefinition of the 
local basis $\{\ev_1(\x),\ev_2(\x)\}$ plays  in fact the same role 
of a gauge transformation, the rotation (\ref{ggt}) producing the 
vector (\ref{ggp}) to transform according to  ${\cal A}_k\rightarrow
{\cal A}_k+\de_k\chi$. The associate geometry-induced magnetic filed 
${\cal B}_k=\ve_{kmn}{\cal B}_{mn}$, ${\cal B}_{mn}=\de_m{\cal A}_n-
\de_n{\cal A}_m$ the {\sl curvature two-form} of ${\cal M}$, may also 
be considered. It is obviously a gauge invariant quantity and, being 
the rotor of a vector field, satisfies the Bianchi identity 
$\mbox{div}\,\Bv=0$.

While the geometry-induced vector potential $\Av$ completely characterizes
the intrinsic geometry of the magnetic line bundle ${\cal M}$, the other
two quantities
\begin{equation}
\begin{array}{l}
{l_1}_k=\ev_1\cdot\de_k\ev_3,\cr
{l_2}_k=\ev_2\cdot\de_k\ev_3,
\end{array}
\end{equation}
describing the flat parallel transport of $R^3$ in the anholonomic frame
$\{\ev_1,\ev_2,\ev_3\}$ may be seen a taking into account its extrinsic
geometry. Since the curvature of the tangent bundle $T\!R^3$ is zero
the three quantities $\Av$, $\Lone$ and $\Ltwo$ are obviously not
independent, being related by the equivalent of the Gauss, Codazzi-Mainardi
and Ricci equations. The ladder, as an example, allows to re-express the 
geometry-induced gauge field $\Bv$ entirely in terms of $\Lone$ and $\Ltwo$ as
\be
\Bv=\Lone\wedge\Ltwo,
\label{Bext}
\ee
${\cal B}_{kl}=({l_1}_k{l_2}_l-{l_1}_l{l_2}_k)/2$, 
$\wedge$ indicating the external product of $R^3$ \cite{F&L80}.
With respect to the point dependent rotation (\ref{ggt}) $\Lone$ and $\Ltwo$
transform as vectors ($\Lone\rightarrow\Lone\cos\chi-\Ltwo\sin\chi$,
$\Ltwo\rightarrow\Lone\sin\chi+\Ltwo\cos\chi$) making the gauge invariance
of $\Bv$ manifest.

\subsection*{Magnetic filed lines geometry}

Thought the geometry of a magnetic filed is completely characterized by two 
independent function (e.g.\ the two independent components of the real magnetic
field $\bv$, or of the geometry-induced magnetic field $\Bv$, etc.\ ) it may
be useful to look at the problem from different points of view. We may wonder, 
as an example, how the intrinsic/extrinsic geometry of the  line bundle 
${\cal M}$ is related to the geometry of magnetic filed lines. To this task
we start by observing that the projection along the field direction
of the two vectors $\Lone$, $\Ltwo$ may be identified  with the two
{\em second fundamental forms} of the embedding of the magnetic field lines in 
$R^3$ \cite{Spi}. In every point of the space the curvature $k$ of the 
magnetic filed line going through that point may so be expressed as
\be
k=\sqrt{(\ev_3\cdot\Lone)^2+(\ev_3\cdot\Ltwo)^2}.
\ee
In a similar way the projection along the filed direction of the 
geometry-induced vector potential $\Av$ have to be identified with
the {\em normal fundamental form} of the embedding of the field lines in $R^3$
(i.e.\ with the connection form induced by the Euclidean geometry
onto the normal bundle of every filed line) \cite{Spi}. Up to the gradient 
of an arbitrary function, representing again the freedom of arbitrarily 
rotating the reference frame in the normal planes, in every point of the
space the torsion $\tau$ of the magnetic field line going through that point
may be written as 
\be
\tau=\ev_3\cdot\Av.
\label{torsion}
\ee
 Curvature and torsion completely characterize the geometry of every 
single magnetic filed line and contain, in principle, all the informations 
relative to the geometry of our problem. On the other hand we may also 
wonder about the global properties of the foliation of $R^3$ in terms
of field lines.
Of particular relevance for the adiabatic motion of a particle 
in an external magnetic field is the possibility of foliating space by
means of surfaces everywhere orthogonal to the field lines. By virtue 
of Frobenius theorem this is controlled by the vanishing of the scalar 
${\cal F}= \ev_3\cdot\,\mbox{rot}\,\ev_3$. In terms of the magnetic line bundle 
geometry
\be
{\cal F}= \ev_1\cdot\Ltwo-\ev_2\cdot\Lone .
\label{Frobenius}
\ee
The magnetic filed lines 
torsion $\tau$ and the Frobenius invariant ${\cal F}$ play a crucial 
role in the description of the anholonomic effects appearing in the 
adiabatic motion of a charged particle in a strong magnetic field.

\section{Adiabatic quantum variables}

We are now ready for the construction of a set of adiabatic operators
adapting to the different time scale behaviours of a quantum particle in a
strong, but otherwise arbitrary, magnetic field. Let us consider therefore
a spin-1/2 particle of mass $m$, charge $e$ and gyromagnetic factor $g$
moving in space under the influence of the {\em inhomogeneous}  magnetic 
field $\bbv(\x)= B\,\bv(\x)$, the physical dimension of the filed being again 
reabsorbed in  the scale factor $B$. Denoting by $\av$ an arbitrary choice of 
the vector  potential, $\mbox{rot}\,\av=\bv$, the dynamics of the system is 
described by the Pauli Hamiltonian
\be
\H/\hbar\omega_B= 
{1\over2}\,\pi_i\pi_i + 
g\, b_i(\x)\sigma_i ,
\label{H_inhom}
\ee
where the kinematical momenta $\pi_i=-i\lb\de_i-a_i(\x)/\lb$ have been 
introduced. The inhomogeneity of the magnetic filed makes 
Hamiltonian (\ref{H_inhom}) to depend on the position operators $\x$,
explicitly through spin term $g\, b_i(\x)\sigma_i$ and implicitly
through the commutation relations of the $\pi_i$s. In spite of the 
simple quadratic dependence of (\ref{H_inhom}) on the kinematical 
momenta, $\pi_1$ and $\pi_2$  are in fact no longer conjugate 
variables and neither commute with $\pi_3$: the set of operators
$\{\pi_i,x^i;i=1,2,3\}$ fulfil the commutation relations
\begin{equation}
\begin{array}{ccc}
[\pi_i,\pi_j] =\, i b_{ij}(\x), \hskip0.7truecm &
[\pi_i,x^j] = -i\lb\delta_i^j,  \hskip0.7truecm &
[x^i  ,x^j] =\, 0,
\end{array}
\label{cr4.1}
\end{equation}
$b_{ij}(\x)=\ve_{ijk}b_k(\x)$ denoting the skew-symmetric two-form associated 
to the field. In the lowest approximation we nevertheless expect the relevant
degree of freedom of the system to be taken into account by the two components
of the  particle's velocity normal to the filed.
Considering the position operators $x^i$s as adiabatic parameters driving 
the fast motion of the system we expect therefore the rapid rotation of 
the particle around its guiding center to be separated from the slow and 
very-slow motion by simply referring the kinematical momenta to the adapted 
frames introduced in the previous section. For the shake of concreteness
we shall indicate by ${R_i}^j(\x)$ the point dependent 
rotation bringing the fixed frame $\{\ex,\ey,\ez\}$ into the adapted frame 
$\{\ev_1(\x),\ev_2(\x),\ev_3(\x)\}$. This allows to decompose the field
$\bv(\x)$ in terms of its norm $b=\sqrt{\bv\cdot\bv}$ and its direction 
$\bv/b={R_i}^j\ez_j$ as  $b_i(\x)=b(\x){R_i}^j(\x)\ez_j$.
 Once the rotation has been performed the kinematical momentum along the 
field direction decouples, up to higher order terms in the adiabatic parameter
$\lb$, from the other two components. The commutator of these, on the other
hand, results to be proportional to $b(\x)$. Stated in a different way, in the 
adapted  frame the particle sees  an effective magnetic filed of constant 
direction and intensity $b(\x)$.
 To make the velocity components normal to the filed in a couple of conjugate 
operators it is now sufficient to rescale them by the point dependent factor 
$b^{-1/2}(\x)$ (see \cite{Ma96}). We shall indicate by ${D_i}^j(\x)$ the 
point dependent dilatation ${D_i}^j=\mbox{diag}(b^{1/2},b^{1/2},1)$ rescaling 
the first and second components of a vector by $b^{1/2}$ and letting the third
one unchanged. 

In order to construct operators adapting to the {\sl fast} time 
scale behaviour of the system two point 
dependent operations have therefore to be performed: (i) a rotation 
${R_i}^j(\x)$ to the local adapted frame and (ii) a dilatation ${D_i}^j(\x)$ 
rescaling the normal components of the kinematical momenta.
 The particle coordinates being not external parameters but dynamical 
variables of the problem these operations will  produce higher
order corrections in the various commutators. We shall therefore proceed
order by order in the adiabatic parameter $\lb$ by constructing sets of 
adiabatic operators fulfilling the desired commutation relation up to
a given order in $\lb$: at the $n$-th order we shall look for a set 
of operators $\{\Pi_i^{(n)}, X^i_{(n)};i=1,2,3\}$ fulfilling the conditions
\begin{itemize}
\item  $\Pi_1^{(n)}$, $\Pi_2^{(n)}$ are a couple of conjugate 
operators up to terms of order $\lb^n$,
\item $\Pi_3^{(n)}$, $X^1_{(n)}$, $X^2_{(n)}$, $X^3_{(n)}$
commute with $\Pi_1^{(n)}$, $\Pi_2^{(n)}$ up to terms of order $\lb^n$,
\item in the limit of a homogeneous filed, $\bv(\x)\rightarrow
\ez$, the adiabatic kinematical momenta $\Pi_i^{(n)}$s and guiding center
operators $X^i_{(n)}$s should reduce to the expressions (\ref{km}) and
(\ref{gco}) respectively.
\end{itemize}
 Our present task being that of separating the fast degree of freedom 
from the slow and very-slow motion, we do not insist for the moment $X^3_{(n)}
-\Pi_3^{(n)}$ and  $X^1_{(n)}-X^2_{(n)}$ to be couple of conjugate operators
as in the homogeneous case.
 
For computational proposes it is very convenient
to use a compact notation which does not distinguish among the physical dislike
directions along and normal to the field. This probably obscures a while
the physical contents of the various expression but greatly simplifies 
formal manipulations. When necessary we will therefore expand
the notation in order to shad light on the physics involved. For the moment
we proceed in the opposite direction by introducing the point dependent
matrix
\be
{\beta_i}^j(\x)={D^{\mbox{\tiny-1}}_i}^k(\x) 
                {R^{\mbox{\tiny-1}}_k}^j(\x)
\label{beta}
\ee
representing the successive application of the two operations necessary
to construct the adapted kinematical momenta in the lowest order. This allows 
to rewrite the skew-symmetric two-form $b_{ij}(\x)$ in terms of $\ve_{kl}=
\ve_{kl3}$ (representing a homogeneous filed directed along $\ez$) 
\be
b_{ij}(\x)={\beta^{\mbox{\tiny-1}}_{\,i}}^k(\x)
           {\beta^{\mbox{\tiny-1}}_{\,j}}^l(\x)\, \ve_{kl}.
\ee
 The matrix ${\beta_i}^j$ and this representation of the filed result to be 
very useful in the construction of the adiabatic quantum variables.

\subsection*{Zero-order operators}
In order to construct the zero-order operators fulfilling the desired
conditions up to terms of order $\lb$ it is sufficient to operate the 
rotation and the dilatation discussed above
\be
\Pz_i={1\over2}\Big\{{\beta_i}^{k},\pi_k\Big\} \label{P0}
\ee
the matrix ${\beta_i}^k$ being evaluated in ${\vec\Xz}\equiv\x$. The
anticommutator $\{\;,\;\}$ is obviously introduced in order to make 
the $\Pz_i$s  Hermitian. A rapid computation confirms our deductions
yielding the commutation relations fulfilled by the zero-order adiabatic 
operators as 
\be
& &\Big[\Pz_i,\Pz_j\Big] =\, i\,\ve_{ij}
 -i\,{\lb\over2}\ve_{ijh}\ve^{hkl}
    \left\{{\beta_k}^m\Gamma_{ml}^{\;\;\;n}, \Pz_n\right\},\nn\\
& &\Big[\Pz_i,\Xz^j\Big] = -i\,\lb\,{\beta_i}^j,         \label{cr4.2}\\
& &\Big[\Xz^i,\Xz^j\Big] =\, 0,                                    \nn
\ee
where $\Gamma_{ki}^{\;\;\;j}=(\de_k{\beta_i}^h)
               {\beta^{\mbox{\tiny-1}}_{\,h}}^j$  and all the functions
are evaluated in ${\vec\Xz}$. $\Pz_1$ and $\Pz_2$ are conjugate operators 
up to ${\cal O}(\lb)$. The commutators depend on the derivative of magnetic 
filed through the vector-valued matrix
\be
(\Gamma_{k})_{i}^{\;\;j}= 
\pmatrix{\ds-{1\over2}{\de_k b\over b} & -{\cal A}_k & - b^{-1/2}\,{l_1}_k \cr
         {\cal A}_k  &\ds -{1\over2}{\de_k b\over b} & - b^{-1/2}\,{l_2}_k \cr
         b^{1/2}\,{l_1}_k & b^{1/2}\,{l_2}_k & 0 }
\label{Gamma}
\ee
allowing to clearly distinguish the effects produced by a variation of
norm of the magnetic filed from that produced by a change of direction.
The ladder are entirely geometrical in character being taken into account 
by the magnetic line bundle connection form $\Av$ and by the two extrinsic 
vectors $\Lone$ and $\Ltwo$.

\subsection*{First-order operators}
Whereas the construction of the zero-order operators is in some way suggested
by the physics of the problem, a more technical effort is required for higher
order terms. The form of the first-oder guiding center operators is
nevertheless  suggested by the corresponding homogeneous expression 
(\ref{gco}),
\be
\Xu^i=\Xz^i+
{\lb\over2}\ve^{kl}\left\{{\beta_k}^i,\Pz_l\right\}, \label{X1}
\ee
the matrix ${\beta_k}^i$ being again evaluated in ${\vec\Xz}$. We 
immediately obtain the new commutation relation 
\be
& &\Big[\Pz_i,\Pz_j\Big] =\, i\,\ve_{ij}
 -i\,{\lb\over2}\ve_{ijh}\ve^{hkl}  
    \left\{{\beta_k}^m\Gamma_{ml}^{\;\;\;n},\Pz_n\right\} +{\cal O}(\lb^2),\nn\\
& &\Big[\Pz_i,\Xu^j\Big] = -i\,\lb\, \delta_i^3 \,{\beta_3}^j+{\cal O}(\lb^2),
                                                         \label{cr4.3}\\
& &\Big[\Xu^i,\Xu^j\Big] = -i\,\lb^2\,\ve^{kl}{\beta_k}^i{\beta_l}^j
                                                         +{\cal O}(\lb^3),  \nn
\ee 
indicating the ${\cal O}(\lb^2)$ decoupling of the adiabatic guiding center 
operators from $\Pz_1$ and $\Pz_2$.
All the functions are now  evaluated in ${\vec\Xu}$. Though
our analysis will be carried out up to ${\cal O}(\lb^2)$, we also wrote
the the first nonvanishing contribution to the commutators
among the $\Xu^i$s, which is of order $\lb^2$. Even if unimportant 
in the present calculation, this allows us to visualize the very-slow 
time scale of the system.

 The construction of the first-order kinematical momenta is performed by
looking for order $\lb$ counterterms to be added to the $\Pz_i$'s. 
These should be  homogeneous second order polynomial in the $\Pz_i$'s with 
coefficients depending on ${\vec\Xu}$. A rather tedious computation 
produces 
\be
\Pu_i=\Pz_i+\lb\,c_{ij}^{klmn}\,
      \Big\{{\beta_m}^h\Gamma_{hn}^{\;\;\;j},
      \Big\{\Pz_k,\Pz_l\Big\}\Big\}, \label{P1}
\ee
where $c_{ij}^{klmn}={1\over24}\ve_{ih}\ve^{kh}
                     (2\delta_j^l+\delta_j^3\delta_3^l)\ve^{mn}+
                     {1\over8}\delta_i^3\ve^{kh}
                     (\delta_j^l+\delta_j^3\delta_3^l)\ve_{hg}\ve^{gmn}$
and all the functions are evaluated in ${\vec\Xu}$.
When expanded these expressions do not look so complicated as a first
sight. We nevertheless insist in keeping this notation which greatly
simplifies the following manipulations. The commutation relations among 
the first-order adiabatic variables are obtained as 
\be
& &\Big[\Pu_i,\Pu_j\Big] =\, i\,\ve_{ij} 
 - i\,{\lb\over4}\ve_{ijk}\ve^{kl}
   \left\{ {\mbox{div}\,\bv\over b}, \Pu_l \right\} +{\cal O}(\lb^2), \nn\\
& &\Big[\Pu_i,\Xu^j\Big] = -i\,\lb\,{\delta_i}^3\,{\beta_3}^j+{\cal O}(\lb^2),
                                                        \label{cr4.4}\\
& &\Big[\Xu^i,\Xu^j\Big] = -i\,\lb^2\,\ve^{kl}{\beta_k}^i{\beta_l}^j 
                                                        +{\cal O}(\lb^3).  \nn
\ee
 It is very interesting to observe that a monopole singularity, that
is a point of nonvanishing divergence, represents an obstruction in 
the construction of the adiabatic operators. Being concerned 
with real magnetic filed we  nevertheless assume $\mbox{div}\,\bv=0$
and carry on in our adiabatic analysis. 
$\Pu_1$ and $\Pu_2$ are then conjugate operators commuting with all the 
remaining variables up to terms of order $\lb^2$ and
the fast degree of freedom decouples from the slow and
very-slow motion up to terms of this order.

\subsection*{A non-canonical set of operators}

 At least in principle it is possible to repeat this construction an arbitrary
number of times getting, as power series in $\lb$, a set of adiabatic  
non-canonical operators $\{\Pi_i,X^i;i=1,2,3\}$ fulfilling the commutation 
relations
\begin{equation}
\begin{array}{ccc}
[\Pi_i,\Pi_j]=\,i\,\ve_{ij},                                 \hskip0.4truecm &
[\Pi_i,X^j]  = -i\,\lb  \,\delta_i^3\,{R^{\mbox{\tiny-1}}_{\,3}}^j,      
                                                             \hskip0.4truecm &
[X^i,X^j]    = -i\,\lb^2\,\ve^{kl}\,b^{-1}\,{R^{\mbox{\tiny-1}}_{\,k}}^i
                                            {R^{\mbox{\tiny-1}}_{\,l}}^j,
\end{array}
\label{cr4.5}
\end{equation}
all the functions being now evaluated in ${\vec X}$. These formal series 
are in general---and have to be \cite{Be87}---not convergent, representing
anyway a very useful tool in the discussion of the adiabatic behaviour of
the system. The description of the problem to a given order $n$ in the 
adiabatic parameter $\lb$ requires the knowledge of the first $n+1$ terms of 
the adiabatic series, so that up to terms of order $\lb^2$ we may identify 
the $\Pi_i$s and $X^i$s with the $\Pu_i$s and $\Xu^i$s respectively. 
An outlook to the commutation relation (\ref{cr4.5}) allows to clearly 
identify the dependence of the canonical structure on the variation of norm
and direction of the magnetic filed. Whereas a suitable redefinition 
of reference frames in $T\!R^3$ allows to separate the fast 
degree of freedom from the others, the very-slow variables are made into 
a couple of non-conjugate operators by an inhomogeneous intensity while 
a variation of the filed direction even produces the mixing of 
very-slow and slow variables.
The description of these by means of couples of conjugate operators requires 
the introduction of curvilinear coordinates in space \cite{Ga59}, the so 
called {\sl Euler potentials} \cite{St70}. We do not insist further on 
this point for the moment observing that under the action of $\Pi_1$,$\Pi_2$  
and $\Pi_3$, $X^1$, $X^2$, $X^3$ the Hilbert space of the system separates in 
the direct sum of two subspaces describing respectively the rapid rotation 
of the particle and the guiding center motion.

\section{Expanding the Hamiltonian}

The adiabatic operators $\vec\Pi$ and $\vec X$ constructed in  the previous 
section have been introduced in such a way to embody the expected features 
of the motion of a quantum  charged particle in a weakly-inhomogeneous 
magnetic field. Their main advantage lies, in fact, in the very suitable 
form assumed by the Pauli Hamiltonian when rewritten in terms of them. 
To this task we have first to invert the power series expressing $\Pi_i$
and $X^i$ in terms of the operators $\pi_i$s and $x^i$s and,
second, to replace these  in (\ref{H_inhom}). This yields 
the Hamiltonian as a power series in the magnetic length $\lb$,
\be
\H=\H^{(0)}+\lb\H^{(1)}+\lb^2\H^{(2)}+...\,
\label{a-exp}
\ee
allowing the adiabatic separation of the fast degree of freedom from the 
slow/very-slow motion and the evaluation of approximate expressions 
of the spectrum  and of the wave functions of the system. 
In order to get the $\pi_i$s and  $x^i$s in terms of the $\Pi_i$s and 
$X^i$s we first recall  that $X^i=\Xu^i+{\cal O}(\lb^2)$. By rewriting 
$\Xu^i$ in terms of the $\Pz_i$s and $\Xz^i= x^i$s, (\ref{X1}), 
$\Pz_i$ in terms of the $\pi_i$s and $x^i$s, (\ref{P0}), and by solving with 
respect to $x^i$, we then obtain $x^i$ as a function of the $\pi_i$s 
and the $X^i$s, $x^i=x^i(\vec\pi,\vec X)$. This allows to rewrite $\Pz_i$
as a function of the $\pi_i$s and $X^i$s. Recalling finally that $\Pi_i
=\Pu_i+{\cal O}(\lb^2)$ and using (\ref{P1}) we immediately get 
$\Pi_i$ in terms of the  $\pi_i$s and $X^i$s, $\Pi_i=\Pi_i(\vec\pi,
\vec X)$. The inversion of this relation, order by order in $\lb$,
allows to get $\pi_i$ and $x^i$ in terms of the adiabatic operators.
The computation gives
\be
\pi_i&=&{1\over2}\Big\{{\beta^{\mbox{\tiny-1}}}_i^{\,j},\Pi_j\Big\}
      +{\lb\over2}\,
       \mbox{c}_{jh}^{klmn} \Big\{
      {\beta^{\mbox{\tiny-1}}}_i^{\,j}
      \beta_m^{\,o}\Gamma_{on}^{\;\;\;h},
      \Big\{\Pi_k,\Pi_l\Big\}\Big\}+{\cal O}(\lb^2), \label{pi}\\
x^i  &=& X^i -\lb\,\ve^{kl}\beta_k^{\,i}\Pi_l+{\cal O}(\lb^2), \label{x}
\ee
where $\mbox{c}_{ij}^{klmn}={1\over2}\delta_i^n\delta_j^k\ve^{ml}-
2c_{ij}^{klmn}$. As a useful check the commutation relations (\ref{cr4.1}) 
may be reobtained by means of the (\ref{cr4.5}).

 The substitution of (\ref{pi}), (\ref{x}) in the Pauli Hamiltonian 
(\ref{H_inhom}) yields immediately the first two terms of the
adiabatic expansion (\ref{a-exp}), 
\be
\H^{(0)}/\hbar\omega_B&=&\,
{1\over2}{\beta^{\mbox{\tiny-1}}}_i^{\,k}{\beta^{\mbox{\tiny-1}}}_i^{\,l}
\;\Pi_k\Pi_l +g\,b_i\,\sigma_i \label{H0}\\
\H^{(1)}/\hbar\omega_B&=&
{\beta^{\mbox{\tiny-1}}}_i^{\,p}{\beta^{\mbox{\tiny-1}}}_i^{\,q}
{\tilde c}_{pj}^{klmn}
\beta_m^{\,o}\Gamma_{on}^{\;\;\;j} 
\Big\{\Pi_k\Pi_q\Pi_l\Big\}
 -g\,\ve^{kl}\,\beta_k^{\,h}\,(\de_hb_i)\,\sigma_i\, \Pi_l \label{H1}\\
... & & \nn
\ee
where the notation $\Big\{\Pi_k\Pi_q\Pi_l\Big\}= \Pi_k\Pi_q\Pi_l + 
\Pi_l\Pi_q\Pi_k$ has been introduced. 
In order to get some more physical insight in this expressions we now 
abandon our compact notation in favour of a more transparent one. 
By recalling the definition (\ref{beta}) of ${\beta_i}^j(\x)$, (\ref{Gamma})
of and $\Gamma_{ij}^{\;\;\;k}(\x)$ and the explicit expression of the 
inhomogeneous dilatation ${D_i}^j(\x)= \mbox{diag}(b^{1/2}(\x),b^{1/2}(\x),1)$,
we rewrite everything in terms of the magnetic field and of other quantities 
capable of a direct physical interpretation. The full expansion of the zero
order Hamiltonian (\ref{H0}) gives
\be
\H^{(0)}/\hbar\omega_B=\, {1\over2}\,{\Pi_3}^2+ b\, \left[J 
                          +g\, (\ev_3\cdot\sigmav)\right],
\label{H0x}
\ee
where $J$ represents the harmonic oscillator Hamiltonian constructed by means
of the canonical variables $\Pi_1$ and $\Pi_2$, $J=({\Pi_1}^2+{\Pi_2}^2)/2$,
and the norm of the magnetic field $b(\vec X)$ is evaluated in the 
adiabatic guiding center operators $\vec X$.
 We observe that while the $\Pi_1$-$\Pi_2$ degree of freedom decouples to 
this order from the slow and very-slow variables the spin does not. 
The separation, up to higher order terms, of the fast motion 
(rotation + spin) requires in fact a subsidiary zero order transformation
which we will perform in the next section.
For the moment let us observe that, up to the spin term, the zero order 
Hamiltonian (\ref{H0x}) precisely embodies the expected behaviour of the 
system: the canonical couple of operators $\Pi_1$-$\Pi_2$ takes into account 
the {\sl fast} rotation of the particle  around  its guiding center, while 
the  non-canonical variables $\Pi_3$-$X^3$ describe the slow motion along
the magnetic field lines by means of an effective ``kinetic energy + potential 
energy'' Hamiltonian. The norm of the magnetic filed $b(\vec X)$ plays the role 
of the effective potential.
As long as ${\cal O}(\lb^2)$ terms are ignored the very-slow dynamical 
variables $X^1$-$X^2$ appear only as adiabatic parameters driving the slow 
motion, whereas a more accurate analysis indicates them as taking into 
account the very-slow drift in the directions normal to the field \cite{Ma96}.

More complicated appears the full expression of the first order 
Hamiltonian (\ref{H1}). The replacement of ${\beta_i}^j(\x)$ and 
$\Gamma_{ij}^{\;\;\;k}(\x)$ by means of (\ref{beta}) and (\ref{Gamma})
yields in fact the expression
\be
& &\;\;\;\;\;\H^{(1)}/\hbar\omega_B\,=\,
                 -b^{-1/2}\ve^{\mu\nu}
                  (\ev_\mu\cdot\mbox{\boldmath$\nabla$} b) \;
                  \left[{2\over3}J_\nu 
                   + g\,(\ev_3\cdot\sigmav)\Pi_\nu\right] 
                -{2\over3}\, b^{1/2}({\ev_\mu}\cdot\Av) \;J_\mu  \nn\\
& &+\left[             {1\over2}\big(\ev_1\cdot\Ltwo-\ev_2\cdot\Lone\big)
                      -(\ev_3\cdot\Av)\right]\; J\, \Pi_3  
            +{1\over4}\big(\ev_1\cdot\Ltwo+\ev_2\cdot\Lone\big)\;
                 ({\Pi_1}^2-{\Pi_2}^2)\, \Pi_3              \nn \\
& &-{1\over4}\big(\ev_1\cdot\Lone-\ev_2\cdot\Ltwo\big)\;
             \{\Pi_1,\Pi_2\}\, \Pi_3                     
                + \, b^{-1/2}\Big[(\ev_3\cdot\Ltwo)\,\Pi_1
                        -(\ev_3\cdot\Lone)\,\Pi_2\Big]\,{\Pi_3}^2    \nn\\
& &-gb^{1/2} \ve^{\mu\nu}
                \Big[(\ev_\mu\cdot\Lone)(\ev_1\cdot\sigmav)
               +(\ev_\mu\cdot\Ltwo)(\ev_2\cdot\sigmav)\Big]\;
                                                         \Pi_\nu,
\label{H1x}
\ee
indicating the first order coupling among the various operators. 
The notation $J_\mu={1\over2}\delta^{\alpha\beta}\Pi_\alpha\Pi_\mu\Pi_\beta$
has been introduced and all the functions are evaluated in $\vec X$.
As expected from dimensional considerations $\H^{(1)}$ depends only on the
first order derivatives of the filed. It is nevertheless worthwhile to stress
that the gradient of the magnetic-field-norm,  $\mbox{grad}\,b=
\mbox{\boldmath$\nabla$}b$, appears only in the first term of the right 
hand side of this expression, all the remaining terms depending only on
the quantities $\Av$, $\Lone$ and $\Ltwo$ completely characterizing 
the intrinsic/extrinsic geometry of the magnetic line bundle ${\cal M}$. 
To a large amount, therefore, the complication of this expression is produced 
by the variation of direction of the magnetic field, that is, by the 
nontrivial geometry of ${\cal M}$. It is not yet time to comment on
the structure of $\H^{(1)}$.
First of all, it is in fact necessary to operate a suitable unitary
transformation separating the zero order fast motion from the other 
degrees  of freedom, that is diagonalizing the the spin term 
$\ev_3\cdot\sigmav$. This will  produce a 
modification of the first order term of the adiabatic expansion. 
Secondly, it is possible to drastically simplify the form of 
$\H^{(1)}$ by operating a suitable  first order unitary transformation. 
 The strategy is nothing else than the quantum equivalent
of the so called {\sl averaging transformation} of classical 
mechanics and results of great help in shading light on the physical
content of (\ref{H1x}).

\section{Quantum averaging transformations}

 A well known strategy in dealing with the adiabatic separation of 
fast and slow variables in classical mechanics consists in performing 
a series of successive canonical transformations (the {\sl averaging
transformations}) separating, order by order in some adiabatic 
parameter, the rapid oscillation of the system from its slow
averaged motion. The analysis depending essentially on the canonical 
structure of the problem  generalizes immediately to quantum 
mechanics, the canonical transformations being replaced by
suitable unitary transformations. The full adiabatic expansion describing 
the motion of a spin degree of freedom adiabatically driven by external 
parameters has been obtained along these lines by M.V.\ Berry \cite{Be87} 
while R.G.\ Littlejohn and S.\ Weigert \cite{L&W93} employed the method 
in discussing the first adiabatic corrections to the semiclassical motion 
of a neutral spinning particle in an inhomogeneous magnetic field.
We shall consider therefore a set of unitary operators 
\be
U^{(n)}=\exp\left\{i\lb^n{\cal L}^{(n)}\right\},
\ee
$n=0,1,...$ such that fast and slow/very-slow degrees of freedom separate
up to ${\cal O}(\lb^{n+1})$ in the Hamiltonian obtained by the 
successive application of $U^{(0)}$, $U^{(1)}$, ...,$U^{(n)}$.
Whereas in classical mechanics it is natural to consider the averaging
transformation as defining new canonical variables, in quantum mechanics 
it appears more convenient to keep the canonical operators fixed and  
transform the Hamiltonian.

\subsection*{Zero-order transformation}

The zero order separation of the fast and slow/very-slow motion
requires the diagonalization of the spin term $gb({\vec X})
(\ev_3({\vec X})\cdot\sigmav)$ of Hamiltonian (\ref{H0x}). 
 Denoting by ${\rho_i}^j(\x)$ the infinitesimal generator
of the rotation ${R_i}^j(\x)$ bringing the fixed frame 
$\{\ex,\ey,\ez\}$ into the adapted frame $\{\ev_1(\x),\ev_2(\x),\ev_3(\x)\}$,
${R_i}^j={(\mbox{e}^\rho)_i}^j\equiv\,\delta_i^j+{\rho_i}^j+
{1\over2}{\rho_i}^k{\rho_k}^j+...$, the aim is achieved by choosing
\be
{\cal L}^{(0)}=-{1\over2}\ve^{ijk}\rho_{ij}(\vec X)\sigma_k,
\label{L0x}
\ee
the matrix $\rho_{ij}={\rho_i}^j$ being evaluated in the guiding center
operators $\vec X$. Because the commutation relations (\ref{cr4.5})
the operator $U^{(0)}$ commute with $\Pi_1$, $\Pi_2$ and therefore with $J$, 
produces ${\cal O}(\lb)$ terms when commuting with $\Pi_3$ 
and ${\cal O}(\lb^2)$ terms when commuting with functions of $\vec X$. 
In evaluating the new Hamiltonian $\H'=U^{(0)}\H {U^{(0)}}^\dagger
={\H^{(0)}}'+\lb{\H^{(1)}}'+...\ $ up to terms of order $\lb^2$
we have therefore to worry only about the action of $U^{(0)}$
on $\sigmav$ and $\Pi_3$. A very rapid computation yields the 
transformation rule
\be
U^{(0)}(\ev_i\cdot\sigmav){U^{(0)}}^\dagger= \sigma_i  +{\cal O}(\lb^2)
\label{U0sigma}
\ee
while the action of $U^{(0)}$ on $\Pi_3$,
$U^{(0)}\Pi_3{U^{(0)}}^\dagger=\Pi_3+U^{(0)}[\Pi_3,{U^{(0)}}^\dagger]$,
may be easily evaluated by computing the commutator in the original 
set of operators $\pi_i$s and $x^i$s and transforming back to adiabatic 
variables
\be
U^{(0)}\Pi_3{U^{(0)}}^\dagger= \Pi_3 + \lb(\ev_3\cdot\Ltwo)\sigma_1
                                     - \lb(\ev_3\cdot\Lone)\sigma_2
                                     + \lb(\ev_3\cdot\Av)  \sigma_3
                                     + {\cal O}(\lb^2).
\label{U0pi3}
\ee
 Subjecting $\H^{(0)}$ and $\H^{(1)}$ to the zero order averaging
transformation $U^{(0)}$ and by using (\ref{U0sigma}) and (\ref{U0pi3})
we obtain the new adiabatic expansion
\be
& &\;\;\;\;\;{\H^{(0)}}'/\hbar\omega_B=\, {1\over2}\,{\Pi_3}^2+ b\, 
                       \left(J +g\, \sigma_3\right), \label{H0'x} \\
& &\;\;\;\;\;{\H^{(1)}}'/\hbar\omega_B\,=\,
                 -b^{-1/2}\ve^{\mu\nu}
                  (\ev_\mu\cdot\mbox{\boldmath$\nabla$} b) \;
                  \left({2\over3}J_\nu + g\,\sigma_3\Pi_\nu\right)
                -{2\over3}\, b^{1/2}({\ev_\mu}\cdot\Av) \;J_\mu  \nn\\
& &+\left[{1\over2}\big(\ev_1\cdot\Ltwo-\ev_2\cdot\Lone\big)J
         -(\ev_3\cdot\Av)(J-\sigma_3)\right]\, \Pi_3  
              +{1\over4}\big(\ev_1\cdot\Ltwo+\ev_2\cdot\Lone\big)\;
                 ({\Pi_1}^2-{\Pi_2}^2)\, \Pi_3              \nn \\
& &-{1\over4}\big(\ev_1\cdot\Lone-\ev_2\cdot\Ltwo\big)\;
             \{\Pi_1,\Pi_2\}\, \Pi_3                     
                + \, b^{-1/2}\Big[(\ev_3\cdot\Ltwo)\,\Pi_1
                        -(\ev_3\cdot\Lone)\,\Pi_2\Big]\,{\Pi_3}^2    \nn\\
& &+            \Big[(\ev_3\cdot\Ltwo)\sigma_1
                    -(\ev_3\cdot\Lone)\sigma_2\Big]\;  \Pi_3
   -            gb^{1/2} \ve^{\mu\nu}
                \Big[(\ev_\mu\cdot\Lone)\sigma_1
                    +(\ev_\mu\cdot\Ltwo)\sigma_2\Big]\; \Pi_\nu, \label{H1'x}\\
& & \;\;\;\;\;... \nn
\ee                                   
 All the functions are evaluated in $\vec X$.
 The fast and slow/very-slow motion are  separated in this way in the zero 
order term of the adiabatic expansion but not in the first order term.

\subsection*{First-order transformation}

The application of the first order averaging transformation $U^{(1)}$ to 
$\H'$ produces the new Hamiltonian $\H''=U^{(1)}\H'{U^{(1)}}^\dagger={\H^{(0)}}'
+\lb({\H^{(1)}}'+i[{\cal L}^{(1)},{\H^{(0)}}'])+...\ $. It is then possible
to simplify the first order term of the adiabatic expansion 
by choosing ${\cal L}^{(1)}$ in such a way that its commutator with 
${\H^{(0)}}'$ cancels as much terms as possible of ${\H^{(1)}}'$. The analysis 
of the commutation relation involved and a little thought indicates that 
it is possible to annihilate all but not the third term of (\ref{H1'x}) 
by choosing 
\be
& &\;\;\;\;{\cal L}^{(1)}\,=\,
                 -b^{-3/2}
                  (\ev_\mu\cdot\mbox{\boldmath$\nabla$} b) \;
                  \left({2\over3}J_\mu + g\,\sigma_3\Pi_\mu\right)
              +{2\over3}\, b^{-1/2}\ve^{\mu\nu}({\ev_\mu}\cdot\Av) \;J_\nu\nn\\
& &           -{1\over8}b^{-1}\big(\ev_1\cdot\Ltwo+\ev_2\cdot\Lone\big)\;
                 \{\Pi_1,\Pi_2\}\,\Pi_3              
              -{1\over8}b^{-1}\big(\ev_1\cdot\Lone-\ev_2\cdot\Ltwo\big)\;
                 ({\Pi_1}^2-{\Pi_2}^2)\,\Pi_3 \nn\\                    
& &           - \,b^{-3/2}     \Big[(\ev_3\cdot\Ltwo)\,\Pi_2
                                   +(\ev_3\cdot\Lone)\,\Pi_1\Big]\,{\Pi_3}^2   
              + \,g^{-1}b^{-1} \Big[(\ev_3\cdot\Ltwo)\sigma_2
                                   +(\ev_3\cdot\Lone)\sigma_1\Big]\;\Pi_3\nn\\
& &           +{g\over g^2-4}b^{-3/2}
               \Big[(\ev_\mu\cdot\Lone)( 2\sigma_1\delta^{\mu\nu}
                                        -g\sigma_2\ve^{\mu\nu}) 
                   -(\ev_\mu\cdot\Ltwo)( 2\sigma_2\delta^{\mu\nu}
                                        +g\sigma_1\ve^{\mu\nu})\Big]\Pi_\nu 
\label{L1x}
\ee 
 The commutators of the zero order Hamiltonian (\ref{H0'x}) with 
the various terms of ${\cal L}^{(1)}$ yields the terms of 
(\ref{H1'x}) times the imaginary factor $i$, in such a way that they 
cancel in the new adiabatic expansion. All the terms but not the third.
It is in fact  immediate to convince that no operator may be found in 
such a way that its commutator with (\ref{H0'x}) produces a term
proportional to $J\Pi_3$ and $\sigma_3\Pi_3$. The third therm of (\ref{H1'x})
may therefore not be removed from the adiabatic expansion representing
a real first order coupling among fast and slow/very-slow motion and not 
a complication produced by a wrong choice of variables. Its 
relevance in the context of the classical guiding center motion  has 
been first recognized by R.G.\ Littlejohn \cite{Lj}.
It is therefore not a surprise to re-find it in the discussion of the 
quantum guiding center dynamics. The  quantum averaging method 
produces so the adiabatic expansion 
\be
&&{\H^{(0)}}''=\,{\H^{(0)}}'\\
&&{\H^{(1)}}''/\hbar\omega_B=
             \,\left[{1\over2}\big(\ev_1\cdot\Ltwo-\ev_2\cdot\Lone\big)J
                -(\ev_3\cdot\Av)(J-\sigma_3)\right]\, \Pi_3,\label{H1''x}\\
&& \;\;\; ...\ .  \nn
\ee
We observe that whereas the zero order terms (\ref{H0'x}) depends 
only on the magnetic-filed-norm $b$ (other than on the commutation 
relations (\ref{cr4.5})) the first order term (\ref{H1''x}) is completely 
characterized by the Frobenius invariant (\ref{Frobenius}),
and by the magnetic filed lines torsion (\ref{torsion}).

\section[]{Quantum guiding center dynamics and \\ 
           magnetic-induced geometric-magnetism}

 The construction of a suitable set of non-canonical operators
embodying the classically expected features of the motion of a charged 
particle in an inhomogeneous magnetic field and the quantum averaging 
method allow us to rewrite the Pauli Hamiltonian (\ref{H_inhom}) 
in such a way that the fast degree 
of freedom---corresponding to the classical rotation of the particle 
around its guiding center---and the spin degree of freedom separate, up 
to terms of order $\lb^2$, from the guiding 
center dynamics. The transformation to the adiabatic operators $\Pi_i$s, 
$X^i$s, (\ref{X1}) and (\ref{P1}), and application of the zero and 
first order quantum averaging operators, (\ref{L0x}) and (\ref{L1x}),
produces in fact the Hamiltonian
\be
\H/\hbar\omega_B=\,{1\over2}\,{\Pi_3}^2+ b\,\left(J +g\, \sigma_3\right) 
            -\lb\left[\tau\,(J-\sigma_3)-{1\over2}{\cal F}\,J \right]\,\Pi_3 
            +{\cal O}(\lb^2).
\label{Heffx}
\ee	
 Disregarding terms of order higher than $\lb$ the operators $J$, 
representing the magnetic moment of gyration of the particle, 
and $\sigma_3$ are constant of motion of the system. 
Frozen the particle in one of its $J$ and $\sigma_3$ eigenstates 
Hamiltonian (\ref{Heffx}) describes therefore the corresponding 
guiding centre dynamics. 
As long as ${\cal O}(\lb^2)$ are ignored $X^1$ and $X^2$ appear
as non-dynamical external adiabatic parameters and only the 
$\Pi_3$-$X^3$ degree of freedom, representing in the classical 
limit the drift of the particle along the magnetic field lines, 
is dynamically relevant.
 To this order, therefore, the quantum guiding center dynamics is 
described by a one  degree of freedom  Hamiltonian given by the sum of 
the kinetic energy  ${\Pi_3}^2/2$ and of an  effective potential proportional 
to $b(\vec X)$. 
$\Pi_3$ being a {\sl slow} variable, that is of the same magnitude of the first 
adiabatic correction, the order $\lb$ term $[\tau(J-\sigma_3)-{\cal F}J/2]
\Pi_3$ may be  identified as a magnetic-like interaction  and reabsorbed 
in the zero order Hamiltonian as a gauge potential. The guiding center 
Hamiltonian rewrites in this way in the familiar form
\be
\H/\hbar\omega_B=\,{1\over2}\,(\Pi_3-\lb\,A(\vec X))^2
   +V(\vec X)
   +{\cal O}(\lb^2),
\label{Hend}
\ee
with
\be
A(\vec X)&=& (J-\sigma_3)\,\tau(\vec X)-{J\over2}\,{\cal F}(\vec X)\label{A},\\
V(\vec X)&=& (J +g\, \sigma_3)\,b(\vec X)\label{V}.
\ee
 As it might be expected from the general discussion of section 3 the 
magnetic filed line torsion $\tau=\ev_3\cdot\Av$ appears 
as (a part of) a gauge potential in the effective slow dynamics,
taking into account the anholonomy produced by the non trivial 
parallel transport of  the magnetic line bundle ${\cal M}$.
Maybe unexpected, at least form this point of view, is the contribution given 
by the Frobenius invariant. Let us in fact 
compare the guiding center motion of a charged particle 
along a magnetic filed line with the propagation of light in a 
coiled optical fiber \cite{C&W86-T&C86} or to the motion of an electron 
constrained on a twisted line \cite{T&T92}. In both  cases---sharing
the same mathematical background---the adiabatic dynamics 
is coupled with an effective vector potential proportional to the torsion of 
the optical fiber or of the twisted line. The  analogue contribution appears
in the guiding center motion, $(J-\sigma_3)\,\tau(\vec X)$, 
but it is not the whole story.
The particle being not homogeneously constrained  in the neighborhood 
of a line,  it results  sensible to the variation of the geometry in
the magnetic 
field lines surrounding the one on which the guiding center si located. 
If all the field lines would have the same geometry,
the foliation of $R^3$ in terms of them would be trivial, the Frobenius
invariant zero and the situation analogue the examples above.
The geometry of this foliation being in general non trivial it yields 
a further contribution to the gauge potential $A(\vec X)$ proportional 
to the Frobenius invariant, $J{\cal F}(\vec X)/2$.
 It is obviously not possible, in the general case, to remove the 
gauge potential (\ref{A}) by means of a suitable choice of gauge.

 In order to make the identification of (\ref{Hend}) with a one---two, 
if we want to consider $X^1$, $X^2$ as dynamical---degree of freedom
Hamiltonian complete it is necessary to replace the $\Pi_i$s, $X^i$s
by a set of canonical operators. The task is achieved by introducing a 
Darboux coordinate frame $\mbox{x}^i=\mbox{x}^i(\x)$, $i=1,2,3$,
bringing the closed skew symmetric two-form $b_{ij}(\x)$ in the
canonical form,
\be
b_{kl}(\x){\de x^k\over\de\mbox{x}^i}{\de x^l\over\de\mbox{x}^j}=\,\ve_{ij}
\ee
($\mbox{x}^1(\x)$, $\mbox{x}^2(\x)$ may be identified with a pair of Euler 
potentials \cite{St70}
for the magnetic filed $\bv(\x)$, $\bnabla\mbox{x}^1\wedge\bnabla\mbox{x}^2
=\bv$, while $\mbox{x}^3(\x)$ with the arc length of the magnetic filed 
lines). Defining $\mbox{X}^i=\mbox{x}^i(\vec X)$, $i=1,2,3$ we get the 
canonical commutation relations $[\Pi_3,\mbox{X}^i]=-i\lb\delta_3^i$ and
$[\mbox{X}^i,\mbox{X}^j]=-i\lb^2\ve^{ij}$ allowing the identification of 
the operators describing the slow and the very-slow degrees of freedom.
 It is in principle possible to start from the beginning by introducing
such curvilinear coordinates in $R^3$ and work out the problem by using
a canonical set of operators \cite{Ga59}.\footnote{The introduction of 
Darboux coordinate 
would produce automatically the framing of $T_{\x}R^3$ by means of an
adapted frame $\{\ev_1(\x),\ev_2(\x),\ev_3(\x)\}$. Our method, on the
other hand, consists in adapting the frame of $T_{\x}R^3$ without
introducing the curvilinear coordinates.}
Nevertheless, whereas the existence of a Darboux coordinate frame is 
always  guaranteed by Darboux theorem, it is hardly ever possible to find it
explicitly and to proceed to the construction of the $\mbox{X}^i$s.
 For this reason---thought the $\Pi_i$s, $\mbox{X}^i$s  appear as the most
natural  variables for the problem---the explicit construction 
of a set of non canonical operators appears as a better 
strategy. 

\section{Conclusions}

 The main difficulty in addressing the separation of fast and slow 
degrees of freedom in the study of an adiabatic system consist 
generally in finding out a suitable set of variables adapting with 
sufficient accuracy to the different time scale behaviours of the 
system. Starting from the homogeneous case and the 
canonical commutation relations (\ref{cr2.2}) we showed
how the analysis of the canonical structure of a charged spinning
particle moving in an external inhomogeneous magnetic field 
leads naturally to the construction---as power series in the
magnetic length $\l_B$---of a suitable set of non-canonical operators
allowing to systematically take into account the coupling between 
spatial and spin degrees of freedom. The new variables fulfil
the very compact commutation relations (\ref{cr4.5}) clearly displaying the 
dependence of the canonical structure on the norm and direction of the 
external magnetic field. In terms of the new operators the Pauli Hamiltonian
rewrites as a power series in the adiabatic parameter $\l_B$ which
may be brought in a particular simple form by operating  suitable 
unitary transformations. In this way the {\sl fast} degree of freedom 
of the  system  representing classically the rapid rotation of the particle
around the guiding center and the spin are separated from the remaining 
degrees of freedom up to terms of order $\l_B^2$. The resulting effective 
guiding center dynamics displays geometric-magnetism:
the coupling with the geometry induced gauge potential (\ref{A}), depending 
on the magnetic filed lines torsion (\ref{torsion}) and on the Frobenius 
invariant (\ref{Frobenius}), and with the scalar potential (\ref{V}), 
proportional to the magnetic field norm. This completely extend to the 
quantum domain 
the previous classical treatments of the problem showing that the anholonomy 
first studied by R.G.\ Littlejohn in the classical guiding center theory
plays an equivalent role in the discussion of the quantum problem. It is 
a feature of the canonical structure of the system after all. The 
geometrical mechanism responsible for the appearance of induced gauge
structures has also been analyzed in some detail and formalized in the 
geometry of the magnetic line bundle ${\cal M}$.

In concluding we observe that our discussion gives in some sense the 
solution of only  half of the problem. The guiding center dynamics is still
characterized by the presence of a {\sl fast} and a {\sl slow} time scale
({\sl slow} $\rightarrow$ {\sl fast}, {\sl very-slow} $\rightarrow$ 
{\sl slow}) and is therefore amenable to a treatment by means of adiabatic 
techniques. Nevertheless, the remaining problem is  not of a 
so deep geometrical nature as the original one and is probably not 
capable of a treatment in general terms.

\section*{Acknoledgments}
 It is a genuine pleasure to thank M.V.\ Berry for hospitality
and a very stimulating ``few days discussion about {\sl fast} and {\sl slow}''
in Bristol. I am also indebted with J.H.\ Hannay, E.\ Onofri and 
J.M.\ Robbins for very helpful discussions on related topics.

\newpage

\end{document}